\documentclass[dvips]{article}
\usepackage{icrctc07}

\title{Comparison of hybrid and pure Monte Carlo shower generators on an event by event basis}
\shorttitle{Hybrid Monte Carlo simulators comparisons}
\authors{J.  Allen$^{1}$, H.-J.  Drescher$^{2}$, G.  Farrar$^{1}$.}
\shortauthors{G.  Farrar and et al}
\afiliations{$^{1}$ Center for Cosmology and Particle Physics, New York University, 4 Washington Place, New York, New York 10003, United States of America\\ $^{2}$ Frankfurt Institute for Advanced Studies, Johann Wolfgang Goethe-Universit\"at, Max-von-Laue Str. 1, 60438 Frankfurt am Main, Germany}
\email{jda292@nyu.edu}

\abstract{\textsc{SENECA} is a hybrid air shower simulation written by H.  Drescher that utilizes both Monte Carlo simulation and cascade equations.  By using the cascade equations only in the high energy portion of the shower, where the shower is inherently one-dimensional, \textsc{SENECA} is able to utilize the advantages in speed from the cascade equations yet still produce complete, three dimensional particle distributions at ground level which capture the shower to shower variations coming from the early interactions.  We present a comparison, on an event by event basis, of \textsc{SENECA} and \textsc{CORSIKA}, a well trusted MC simulation code.  By using the same first interaction in both \textsc{SENECA} and \textsc{CORSIKA}, the effect of the cascade equations can be studied within a single shower, rather than averaged over many showers.  Our study shows that for showers produced in this manner, \textsc{SENECA} agrees with \textsc{CORSIKA} to a very high accuracy with respect to densities, energies, and timing information for individual species of ground-level particles from both iron and proton primaries with energies between 1 EeV and 100 EeV.  Used properly, \textsc{SENECA} produces ground particle distributions virtually indistinguishable from those of \textsc{CORSIKA} in a fraction of the time.  For example, for a shower induced by a 10 EeV proton, \textsc{SENECA} is 10 times faster than \textsc{CORSIKA}, with comparable accuracy.}

\begin{document}
\maketitle

\section{Introduction}
\textsc{SENECA} is a hybrid air shower simulation, combining cascade equations with Monte Carlo (MC) to quickly produce fully descriptive ground particle distributions \cite{DFASS}.  High energy cosmic ray experiments, such as the Pierre Auger Observatory, are dependent upon air shower simulations to understand shower development and detector response.  Higher energy simulation has traditionally resulted in a significant computational problem.  Thinning has been used to cut down computation times, but always at the cost of accuracy.  Thinning introduces artificial fluctuations to the lateral distribution function (LDF).

By beginning with a high energy MC stage, using cascade equations in the one-dimensional regime of the shower, and MC elsewhere, one can reproduce the longitudinal profile \cite{OLD} and the LDF \cite{DFASS} with a high degree of accuracy.  For high energy showers, using a hybrid approach can reduce computation times by over a factor of ten \cite{DFASS} and allow for a superior thinning method, as described below.

\subsection{Event by event method}
\textsc{CORSIKA} has been well tested and is well trusted, and is therefore used as a standard to which \textsc{SENECA} can be compared.  Previous studies have already shown that \textsc{SENECA} and \textsc{CORSIKA} agree well when comparing the average LDF of 10 EeV proton primary showers \cite{DFASS}, as well as the average longitudinal profile for primary energies between 1 EeV and 100 EeV \cite{OLD}.  We will present a technique which allows for meaningful comparisons between \textsc{SENECA} and \textsc{CORSIKA} on an event by event basis.

A large contribution to the natural fluctuations of showers with identical primaries comes from the height and dynamics of the first interaction.  By using the same first interaction in both \textsc{SENECA} and \textsc{CORSIKA}, showers can be compared on an individual basis.  This can be done by using the STACKIN option in \textsc{CORSIKA} \cite{CUG}.  The first interaction is simulated using \textsc{SENECA}, and secondary particles are written to a file which may be read in by \textsc{CORSIKA}.  This was done for vertical proton showers of energy 1 EeV, 10 EeV, and 100 EeV at thinning levels of $10^{-6}$, $10^{-7}$, and $10^{-8}$.

In order for this method to give meaningful results, it is crucial to use the same hadronic and electromagnetic models.  In the hadronic case, QGSJet01 and Gheisha 2002 were used as the high energy and low energy models, respectively.  If any discrepancies existed in the hadronic and muonic components, presumably arising from the use of cascade equations, our comparisons would reveal them.  While conducting this study, we found that the photonuclear interactions in SENECA were not described with enough realism.  This leads to discrepancies in the electromagnetic component at the level of 10\% at 1 km and greater than 20\% at 3 km.  In order to make a valid comparison of the electromagnetic component, photonuclear interactions were turned off in both SENECA and CORSIKA.  The problem with the photonuclear treatment in the original Seneca distribution is being corrected and a full comparison will be presented elsewhere.

\begin{figure}
\begin{center}
\includegraphics [width=0.33\textwidth, angle=-90]{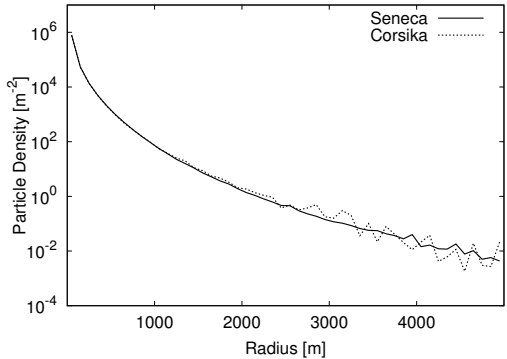}
\end{center}
\caption{Total LDF for a vertical 10 EeV proton primary shower.  Solid line represents \textsc{SENECA}, dashed line represents \textsc{CORSIKA}. The LDF was calculated by adding all the weight in 100 m wide rings.
}\label{figLDFs}
\end{figure}
    
\subsection{Lateral Distribution Function}
\begin{figure}
\begin{center}
\includegraphics [width=0.55\textwidth, angle=-90]{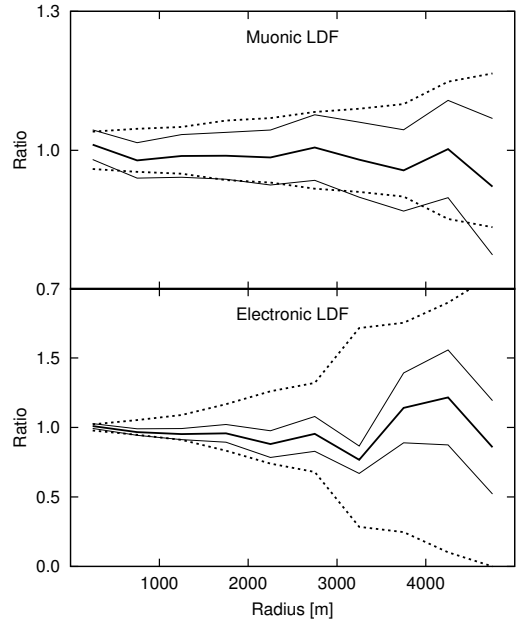}
\end{center}
\caption{Ratio of \textsc{SENECA} LDF to \textsc{CORSIKA} LDF, for muons and electrons.  One-sigma bands are shown for \textsc{CORSIKA} by dashed lines, and for \textsc{SENECA} by the thinner solid lines.  These errors include artificial and natural fluctuations.  To create the LDFs that are compared, an average of 50 showers was used, all with the same first interaction.
}\label{figRatio}
\end{figure}

It was found that the results of our study are independent of the energy of the primary, so our discussion will focus on 10 EeV primary showers.  By eye, the LDF for \textsc{SENECA} and \textsc{CORSIKA} are in agreement.  Figure \ref{figLDFs} shows the lateral density of particles for a single \textsc{SENECA} and \textsc{CORSIKA} shower.  Discrepancies do not arise until a radius where fluctuations become large.  In order to make a robust comparison, a library of 50 showers, all with the same first interaction, was generated, for both \textsc{CORSIKA} and \textsc{SENECA}, using a thinning level of $10^{-7}$.  The results of this comparison are shown in figure \ref{figRatio}.  Discrepancies in the averaged density of particles are less than 5\% out to a radius of 3km, for both hadrons and muons, and the averaged LDFs are in agreement within one sigma at all radii.

\subsection{Energy Distribution}
\begin{figure}
\begin{center}
\includegraphics [width=0.55\textwidth, angle=-90]{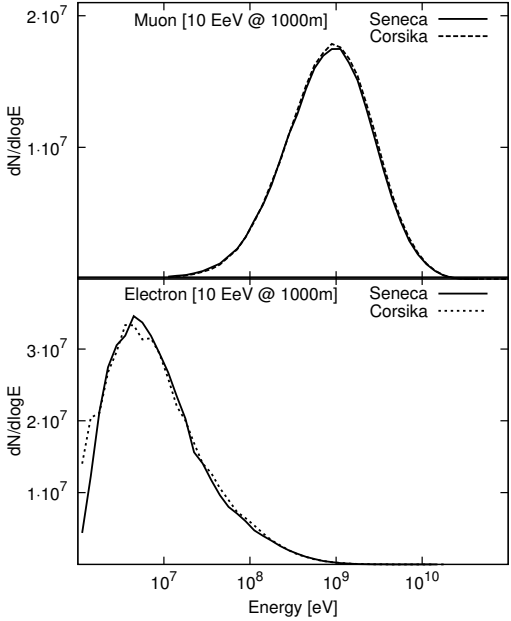}
\end{center}
\caption{Energy distribution of muons and electrons at a radius of 1000 m, of 50 10 EeV proton induced showers. The first interaction of all the showers is identical. The solid and dashed lines represent \textsc{SENECA} and \textsc{CORSIKA}, respectively.
}\label{figEngDist}
\end{figure}

A comparison of the energy distributions of electrons and muons can be seen in figure \ref{figEngDist}.  Similar to what was found in the LDF comparison, the muonic energy distributions agree nicely.  The average value of the distributions at both radii differ by at most 2\%.  

\begin{figure}
\begin{center}
\includegraphics [width=0.33\textwidth, angle=-90]{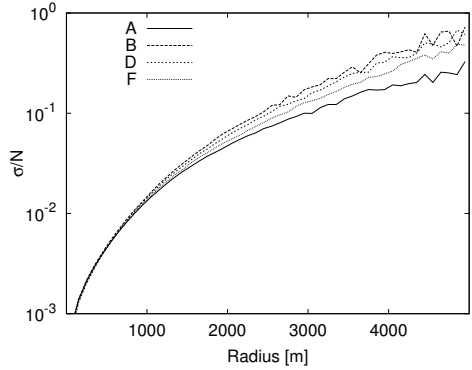}
\end{center}
\caption{An estimate of artificial fluctuations for a variety of thinning methods (See Table \ref{efftab}), as explained in equation (\ref{eqnSOR}).
}\label{figSORs}
\end{figure}

\subsection{Thinning}
The principle motivation for using cascade equations is the significant increase in speed.  However, the thinning methods used by \textsc{SENECA} and \textsc{CORSIKA} are fundamentally different.  In \textsc{SENECA}, the shower is followed exactly until the end of the cascade equations.  Particles are sampled from source functions and assigned a weight that they will keep until the end of the simulation \cite{DFASS}.  The nearest method available in \textsc{CORSIKA} is optimized thinning with maximum weight limits, in which Hillas or statistical thinning is effectively applied over an energy range specified by the parameters of the thinning \cite{WLT}.

\begin{table*}
\begin{center}
  \begin{tabular}{| c | c | c | c | c | r |}
    \hline
    \multicolumn{6}{|c|}{Efficiency Comparison} \\ \hline
    Label & Generator & Thinning Method & $W_{(max)}^{EM}$ & $W_{(max)}^{Had}$ & $Q_A$ \\ \hline
    A & \textsc{SENECA} & - & 10,000 & 1,000 & 1 \\ \hline
    B & \textsc{SENECA} & - & 10,000 & 10,000 & 3.2 \\ \hline
    C & \textsc{SENECA} & - & 4,000 & 4,000 & 3.2 \\ \hline
    D & \textsc{CORSIKA} & Opt.  $10^{-5}$ & 10,000 & 10,000 & 8.7 \\ \hline
    E & \textsc{CORSIKA} & Opt.  $10^{-6}$ & 10,000 & 10,000 & 14.5 \\ \hline
    F & \textsc{CORSIKA} & Opt.  $10^{-5}$ & 1,000 & 10,000 & 7.3 \\
    \hline
  \end{tabular}
  \caption{This table compares the efficiency of various thinning methods, in both \textsc{SENECA} and \textsc{CORSIKA}, to the efficiency of running \textsc{SENECA} with a hadronic weight of 1,000 and an electromagnetic weight of 10,000.  $Q_A$ is as defined in equation (\ref{eqnQ}), where thinning method A has been used as the reference.  So thinning A is 3.2 times more efficient than thinning B.  $Q_A$ for thinning method A is 1 by definition.  In the case of \textsc{SENECA}, $W_{(max)}^{EM}$ refers to the final weight of electromagnetic particles, whereas for \textsc{CORSIKA} it refers to the maximum allowable weight, and likewise for $W_{(max)}^{Had}$.}
  \label{efftab}
  \end{center}
\end{table*}

A comparison of the artificial fluctuations induced by thinning is shown in figure \ref{figSORs}.  $\frac{\sigma}{N}$ is defined as
\begin{equation}
\frac{\sigma}{N} = \frac{\sqrt{\sum_i W_i^2}}{\sum_i W_i} ,
\label{eqnSOR}
\end{equation}
where the sum is over all tracked particles and $W_i$ is the weight of an individual particle. $\sigma$ provides a measure of the artificial fluctuations induced by thinning. As can be seen, even when the maximum weights used are identical for \textsc{CORSIKA} and \textsc{SENECA}, \textsc{SENECA} produces less artificial fluctuations.  This is because the cascade equations can be used to follow the shower exactly to quite low energies, below which there are so many particles that the fluctuations from thinning are small.  In the example shown in figure \ref{figSORs}, cascade equations were used down to $10^{4}$ GeV and 10 GeV for the hadronic and EM portions, respectively.  The \textsc{SENECA} shower has artificial fluctuations similar to a \textsc{CORSIKA} shower using thinning levels $10^{-6}$ and $10^{-9}$ for the hadronic and EM portions, respectively.  The cascade equations naturally lend themselves to very efficient and effective thinning.

\subsection{Efficiency}
The relative efficiency of a thinning method and simulation, which takes into account speed and minimization of artificial fluctuations, can be defined as follows \cite{WLT}
\begin{equation}
Q_A = \left(\frac{\sigma}{\sigma_A}\right)^{2}\left(\frac{t}{t_A}\right)
\label{eqnQ}
\end{equation}
where $t$ is the computation times of the simulation, and $\sigma$ is defined as the numerator of the right hand side of equation (\ref{eqnSOR}). $\sigma_A$ and $t_A$ correspond to thinning method A in table \ref{efftab}, which is used as a reference.  For example, thinning method D has a $Q_A$ value of 8.7, thus, thinning method A is 8.7 times more efficient than D.  Table \ref{efftab} compares a variety of thinning methods, in both \textsc{SENECA} and \textsc{CORSIKA}.  Thinning method A was found to be the most efficient method for a 10 EeV shower.

\subsection{Conclusion}
The hybrid simulation method employed by \textsc{SENECA} has two very significant advantages over a pure MC simulation: speed and accuracy.  \textsc{SENECA} is at least 7.3 times more efficient than \textsc{CORSIKA}, in terms of quickly producing showers with minimized artificial fluctuations.  As has been demonstrated, this increase in efficiency comes at no cost in physical accuracy.  \textsc{SENECA} is able to produce particle densities and particle energy distributions that are consistent to those produced by \textsc{CORSIKA}, when the interaction models used in the simulation are the same.  Further details of the comparisons and the results of the analysis with photonuclear interactions restored will be presented elsewhere.

\bibliography{icrc1278}
\bibliographystyle{plain}

\end{document}